\documentclass[fleqn,11pt]{SelfArx} 

\usepackage[english]{babel} 

\usepackage{lipsum} 

\usepackage{natbib}

\usepackage{tabularx}

\definecolor{color1}{RGB}{0,0,90} 
\definecolor{color2}{RGB}{0,20,20} 

\usepackage{hyperref} 

\hypersetup{
	hidelinks,
	colorlinks,
	breaklinks=true,
	urlcolor=color2,
	citecolor=color1,
	linkcolor=color1,
	bookmarksopen=false,
	pdftitle={Title},
	pdfauthor={Author},
}

\JournalInfo{Accepted for publication in Solar Physics, 2026}
\Archive{DOI 10.1007/s11207-026-02732-z}

\PaperTitle{Cycle Variation in the Occurrence of Great Soft X-ray Solar Flares}
\PaperSubTitle{Cycle Variation of Great Soft X-ray Solar Flares}

\Authors{V.N.~\ Obridko\textsuperscript{1}, M.M.~\ Katsova\textsuperscript{2}, D.D.~\ Sokoloff\textsuperscript{1,3}*}

\affiliation{\textsuperscript{1}\textit{IZMIRAN, 4, Kaluzhskoe Shosse, Troitsk, Moscow, 108840, Russia}}
\affiliation{\textsuperscript{2}\textit{Sternberg Astronomical Institute, Lomonosov Moscow State University, Moscow, 119234, Russia}}
\affiliation{\textsuperscript{3}\textit{Department of Physics, Lomonosov Moscow State University, Moscow, 119991, Russia}}
\affiliation{*\textbf{Corresponding author}: sokoloff.dd@gmail.com}

\Keywords{Flares --- Relation to magnetic field --- Solar cycle --- Observations}
\newcommand{\keywordname}{Keywords} 

\Abstract{
Powerful nonstationary processes such as flares and coronal mass ejections are detected
mainly near the solar cycle maxima. However, the analysis of the great flares since 1975,
when the GOES monitoring started, leads us to more precise conclusions. We have traced
the latitudinal distribution of X-class flares and their dependence on the phase of the cycle
throughout four cycles. Besides, we compared directly their positions with the magnetic field
structure. It was found that the relatively weak X4--7 flares occur at relatively high latitudes
(15 -- 20 degrees), where two waves of activity converge, one moving toward the equator
and the other, the wave of the following cycle, directed poleward. Such events are observed
almost constantly 1 -- 2 years before the cycle maximum. At the same time, the number of
the powerful X-ray flares increases sharply during the maximum phase, but the greatest
X10 flares are observed at the beginning of the decline phase, where wave interactions still
persist, and throughout the decline phase. Note that they are virtually absent during the cycle
growth phase. Thus, we conclude that the greatest X-ray flares begin to appear 1 -- 2 years
before the maximum number of sunspots in the overlapping phase, when different kinds of
activity waves coexist on the Sun at relatively high latitudes, and, then, continue to appear
at the boundary, which separates the wave of local fields and the poleward wave of the
following cycle.
}

\begin{document}

\maketitle 


\thispagestyle{empty} 


\section*{Introduction} 

\addcontentsline{toc}{section}{Introduction} 

Solar flares are an important part of solar activity, which is very interesting both by itself
and because the flares lead to different phenomena in the immediate Earth surroundings.

Naturally, the most energetic flares attract particular attention. The flares originate during
the evolution of the solar magnetic field and are expected to be somehow associated with the
solar magnetic activity cycle. The point, however, is that this link is far from straightforward.
For example, one cannot expect that the number of flares is just proportional to the magnetic
field strength, because many other factors are important for the formation of flares (e.g., see
\cite{PF02}). The observational information concerning the greatest X-ray solar
flares is collected in particular by \cite{Hetal24}.

Nevertheless, some relation between the flares and the activity cycle seems to exist (e.g.,
\cite{BTetal25}). Of course, it is generally accepted that the most powerful flare should be
the result of large-scale transformation of the magnetic energy to other forms, which always
happens at the decay phase of the solar cycle.

However this idea is too general and, obviously, needs a more detailed analysis and clarification.
In particular, it would be useful to learn in more detail the latitude-time distribution
of the greatest flares in comparison with the shape of the magnetic cycle. However, it is
problematic to realize this idea straightforwardly, because the number density of the greatest
flares is, by definition, quite low and various irregularities in both distributions do not
allow us to learn more than a general statement above.

We propose to apply the method of superposed epochs to obtain a synthetic space-time
distribution of the greatest flares. The superposed epoch analysis consists of two stages.
First, a time interval $t$ from the beginning of the cycle, $t_0$, is selected as the abscissa, 
which is, then, normalized to the length of the cycle, $T$. This makes all cycles comparable with
the equal independent variable $F = (t - t_0)/T$. In the process, the ordinate values do not
change. After that, all figures are combined. In the process of combining, we can choose
the initial point $F = 0$ (as in Figure 3a) or, for example, the phase maximum $F_m$ (as in
Figure 3b) as the zero point of the abscissa axis.

\section{Data}

We are interested in the solar X-ray flares larger than X4 (Tables 1, 2). 
The GOES (Geostationary operational environmental satellite) soft X-ray flare list was used. The flare classes
are determined by their peak fluxes as follows: B stands for fluxes 
$(1 - 9.9) \times 10^{-7} \,\mbox{Wm}^{-2}$,
C stands for $(1 - 9.9) \times 10^{-6} \,\mbox{Wm}^{-2}$, 
M corresponds to $(1 - 9.9) \times 10^{-5} \,\mbox{Wm}^{-2}$, 
and X stands for $(1 - 9.9) \times 10^{-4} \,\mbox{Wm}^{-2}$. 
As is shown in \cite{KOSL2022}, the occurrence rate of the
B and C class flares depends weakly on spottedness, while the occurrence rate of strong 
M and X class flares depends very strongly.

The flares larger than X10 (Table 2) come mainly from 
\url{https://www.spaceweatherlive.com/en/solar-activity/top-50-solar-flares.html}. 
This list involves 12 flares larger than X10. We also add the flares from the lists by 
\cite{Hetal24, BTetal25}.
The list of the flares with magnitudes from X4 to X10 (Table 1), as well as the earlier events
comes mainly from 
\url{https://www.ngdc.noaa.gov/stp/space-weather/solar-data/solar-features/solar-flares/x-rays/goes/xrs}. 
The data on the Halloween Solar Storm flares come from
\url{https://www.solarmonitor.org/index.php}.
The factor 1.43 suggested by \cite{Hetal24} is taken into account. 
The published format of the flare magnitudes is conserved.

We compare the latitude-time distribution of the largest X-ray solar flares with the corresponding
distribution of the surface solar magnetic field based on the data from \cite{H15}.

\begin{table}
\caption{Solar flares with X-ray magnitude between 4 and 10. A- year, B - month, C - day, D - X-ray magnitude of the flare (X-class), E - latitude, F - longitude}
 \begin{tabularx}{\textwidth}{XXXXXXXX}
    \hline
A. B. C	& D &	E	& F  & A. B. C& D & E &F \cr
\hline
1978.04.28&	X7.15 &	22	& -41 &
1979.08.18& X8.58 &	10 & -90 \cr
1979.08.20&	X7.15&	5 &	-76 &
1979.09.20&	X7.15&	6 &	-33 \cr
1980.04.04&	X7.15&	24 &	34 &
1981.04.24&	X8.44&	18 &	50 \cr
1982.06.04&	X8.44&	-10 &	-55 &
1982.11.26&	X6.43&	-11 &	87 \cr
1982.12.15&	X7.15&	-10 &	-15 &
1983.02.03 & X5.86 & -19 & 8 \cr
1984.05.19 & X5.86 & -7 & -67 &
1985.01.22&	X4.7&	-8&	38 \cr
1988.12.16&	X6.72&	26&	-37 &
1989.03.10&	X6.43&	31&	-22 \cr
1989.03.17&	X6.5&	33&	60 &
1989.06.15&	X9.29&	-21&	-8  \cr
1989.10.25&	X8.16&	-30&	57 &
1990.05.21&	X7.86&	35&	36 \cr
1991.10.27&	X9.72&	-13&	-15 &
1998.08.18&	X7.03&	32&	-72\cr
1998.08.19&	X5.57&	32&	-72&
1998.11.22&	X5.37&	-27&	75\cr
1998.11.28&	X4.77&	19&	-44 &
2000.07.14&	X8.21&	17&	-3   \cr
2000.11.26&	X5.83&	22&	21  &
2001.04.6&	X8.08&	-21&	-47 \cr
2001.08.25&	X7.7&	-18&	-42 &
2001.12.13&	X8.9&	14&	-18 \cr
2001.12.28&	X4.99&	-23&	-73&
2002.07.20&	X4.74&	-12&	-68 \cr
2002.07.23&	X6.98&	-12&	-68 &
2002.08.24&	X4.54&	-8&	85\cr
2003.05.28&	X5.17&	-7&	19&
2003.10.23&	X7.77&	-16&	-81\cr
2003.11.3&	X5.61&	8&	68&
2004.07.16&	X5.24&	-10&	-40\cr
2005.01.17&	X5.52&	13&	15&
2005.09.08&	X7.77&	-12&	-83\cr
2005.09.09&	X8.87&	-9&	-67&
2006.12.06&	X9.4&	-6&	-72\cr
2006.12.13&	X4.88&	-6&	21&
2011.08.09&	X9.96&	18&	68\cr
2012.03.07&	X7.79&	17&	-29&
2013.05.14&	X4.64&	11&	-63\cr
2013.11.05&	X4.93&	-9&	-49&
2014.02.25&	X7.13&	-15&	-86\cr
2014.10.24&	X4.58&	-14&	6&
2023.12.31&	X5.01&	5&	-75\cr
2024.02.22&	X6.37&	18&	-26&
2024.05.11&	X5.89&	-18&	67\cr
2024.05.14&	X8.79&	-19&	0&
2024.09.14&	X4.54&	-18&	-48\cr
2024.10.01&	X7.1&	-15&	-17&
2024.10.03&	X9&	-15	&8\cr
2025.11.11&	X5.1&	24&	34&
2026.02.01&	X8.11&	14&	-34\cr
\hline       
	\end{tabularx}
    \label{T1}
\end{table}

\begin{table}
\caption{Solar flares with X-ray magnitude over 10. A- year, B - month, C - day, D - X-ray magnitude of the flare (X-class), E - latitude, F - longitude}
\begin{center}
 \begin{tabularx}{\textwidth}{XXXXXXXX}
    \hline
A.B.C	& D &	E	& F  & A.B.C& D & E &F \cr
\hline
1978.07.11&	X45.9&	20&	-46&
1980.11.06&	X12.8&	-12&	-74\cr
1982.06.03&	X12.8&	-9&	-72&
1982.06.06&	X14.7&	-9&	-25\cr
1982.07.12&	X10.1&	11&	-37&
1982.12.15&	X19&	-10&	-24\cr
1982.12.17&	X14.7&	-8&	21&
1984.04.24&	X19.7&	-12&	-43\cr
1984.05.20&	X14.8&	-9&	-52&
1989.03.06&	X19.3&	35&	-69\cr
1989.08.16&	X28&	-18&	84&
1989.09.29&	X14.2&	-20&	90\cr
1989.10.19&	X23.1&	-27&	-10&
1990.05.24&	X14.1&	33&	78\cr
1991.01.25&	X15.7&	-16&	-78&
1991.03.04&	X10.4&	-20&	-88\cr
1991.03.22&	X13.8&	-26&	-28&
1991.06.01&	X28.3&	25&	-90\cr
1991.06.04&	X32&	30&	-70&
1991.06.06&	X30.2&	33&	-44\cr
1991.06.09&	X15.1&	34&	-4&
1991.06.11&	X17&	31&	17\cr
1991.06.15&	X34.7&	33&	69&
1992.11.02&	X13.3&	-26&	87\cr
1997.11.06&	X13.1&	-18&	63&
2001.04.02&	X29.6&	19&	72\cr
2001.04.15&	X21.1&	-20&	85&
2003.10.28&	X25.7&	-16&	-8\cr
2003.10.29&	X15.5&	-15&	2&
2003.11.02&	X13.3&	-14&	56\cr
2003.11.04&	X43.2&	-19&	83&
2005.01.20&	X10.2&	14&	61\cr
2005.09.07&	X24.6&	-11&	-77&
2006.12.05&	X13.1&	-7&	-68\cr
2011.08.09&	X10.7&	17&	69&
2017.09.06&	X14.8&	-8&	33\cr
2017.09.10&	X13&	-8&	88&
&&&\cr
\hline       
    \end{tabularx}
\end{center}
\label{T2}
\end{table}

\section{Latitude-time distribution of the largest X-ray flares}

In Fig. 1, we represent the latitude-time distribution of the flares listed in  Table 1 (blue points) and Table 2 (red points) on the contour map of the surface magnetic field (upper panel) in comparison with the sunspot data (lower panel). The magnetic field map was constructed by averaging the absolute values of the WSO  (Wilcox solar observatory) measurements over longitude in each rotation (\url{http://wso.stanford.edu}). SSN (Sunspot number data) data were taken from WDC-SILSO (World Data Center - Sunspot Index and long-term solar observations), Royal Observatory of Belgium, Brussels.

We should note that our task does not include the analysis of cyclic variation. The figures only represent data for four full cycles 21-24, i.e., from 1976 to 2017. At the same time, table 1 also provides data on 9 X-ray flares that occurred in cycle 25.

We see that the flares demonstrate a cyclic variation and  prefer the descending branch of the cycles. However, the density of the points is too low to allow an immediate final conclusion. 

\begin{figure}    
\begin{center}
\includegraphics[width=0.9\textwidth]{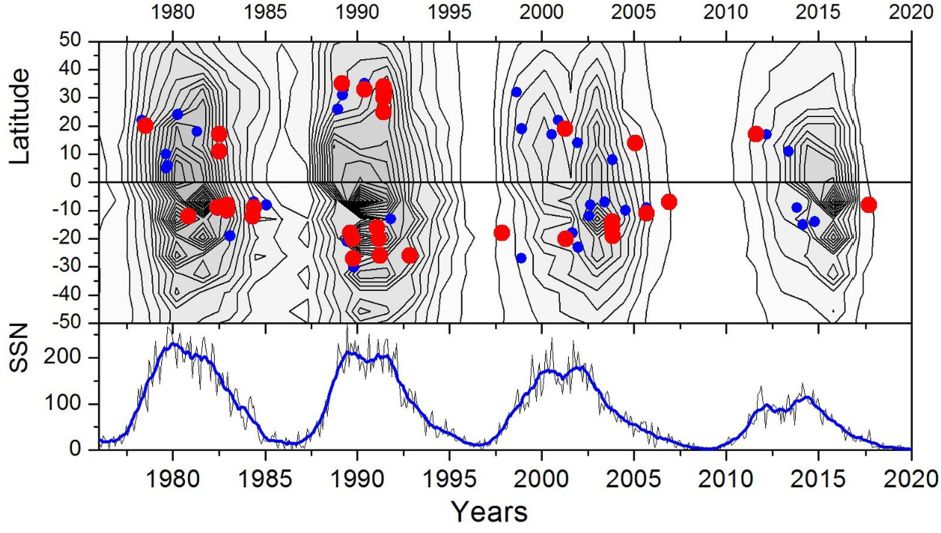}
\small
        \caption{Latitude-time distribution of the largest X-ray solar flares against the background of the surface magnetic field butterfly diagram shown with thin solid lines (upper panel) in comparison with the SSN cyclic variation (lower panel). The red points correspond to the flares of magnitude larger than X10, the blue ones, to the flares of magnitudes between X4 and X10.}
           
\end{center}
\label{F1}
\end{figure}

In Fig. 2, the positions of all flares under discussion are represented on the time-latitude diagram  to show the 22-year variation and the formation of patterns in both hemispheres.

\begin{figure}    
\begin{center}
\includegraphics[width=0.7\textwidth]{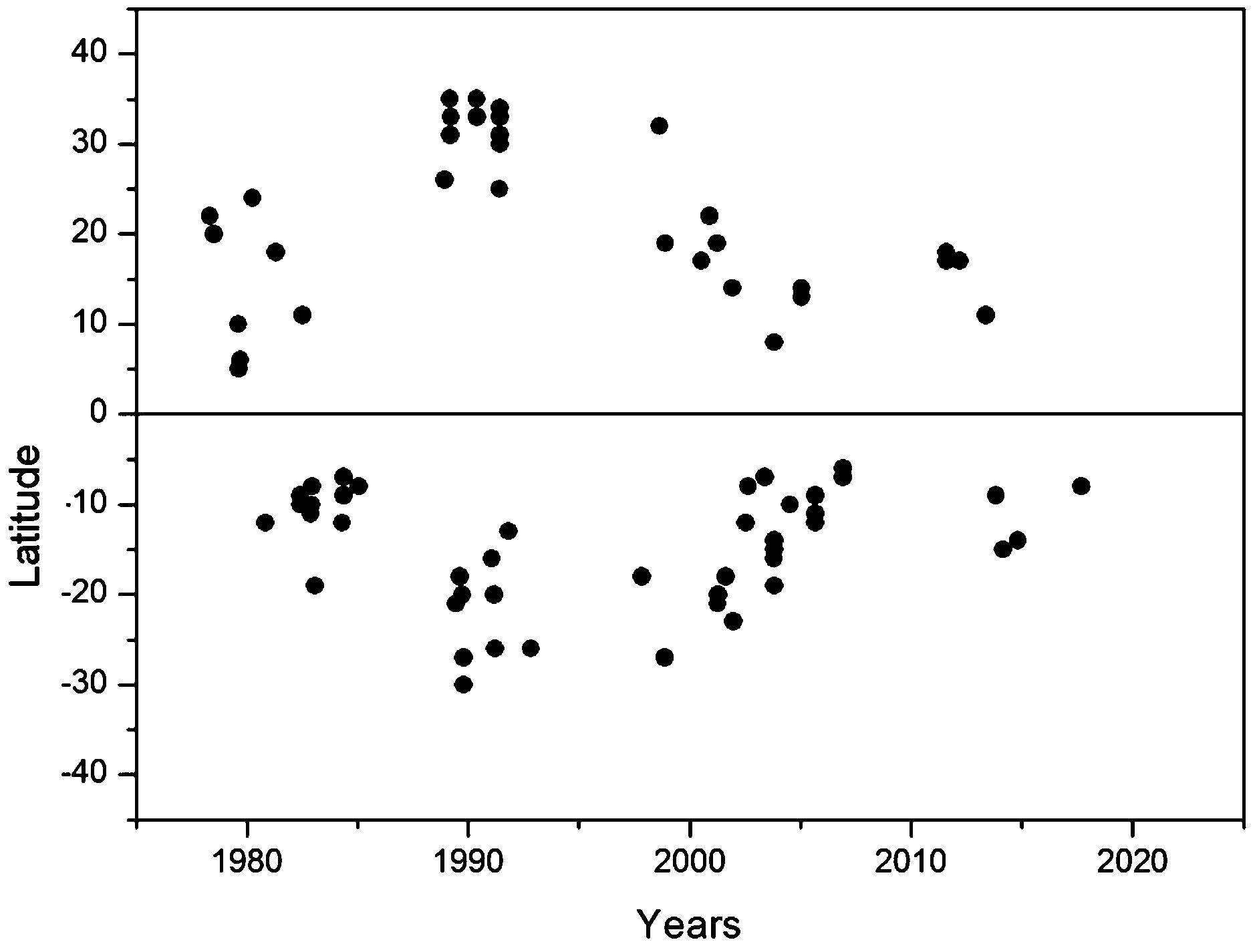}
\small
        \caption{Time-latitude diagram for the flares under investigation.}
\end{center}
\label{F2}
\end{figure}

\begin{figure}    
\begin{center}
\includegraphics[width=0.8\textwidth]{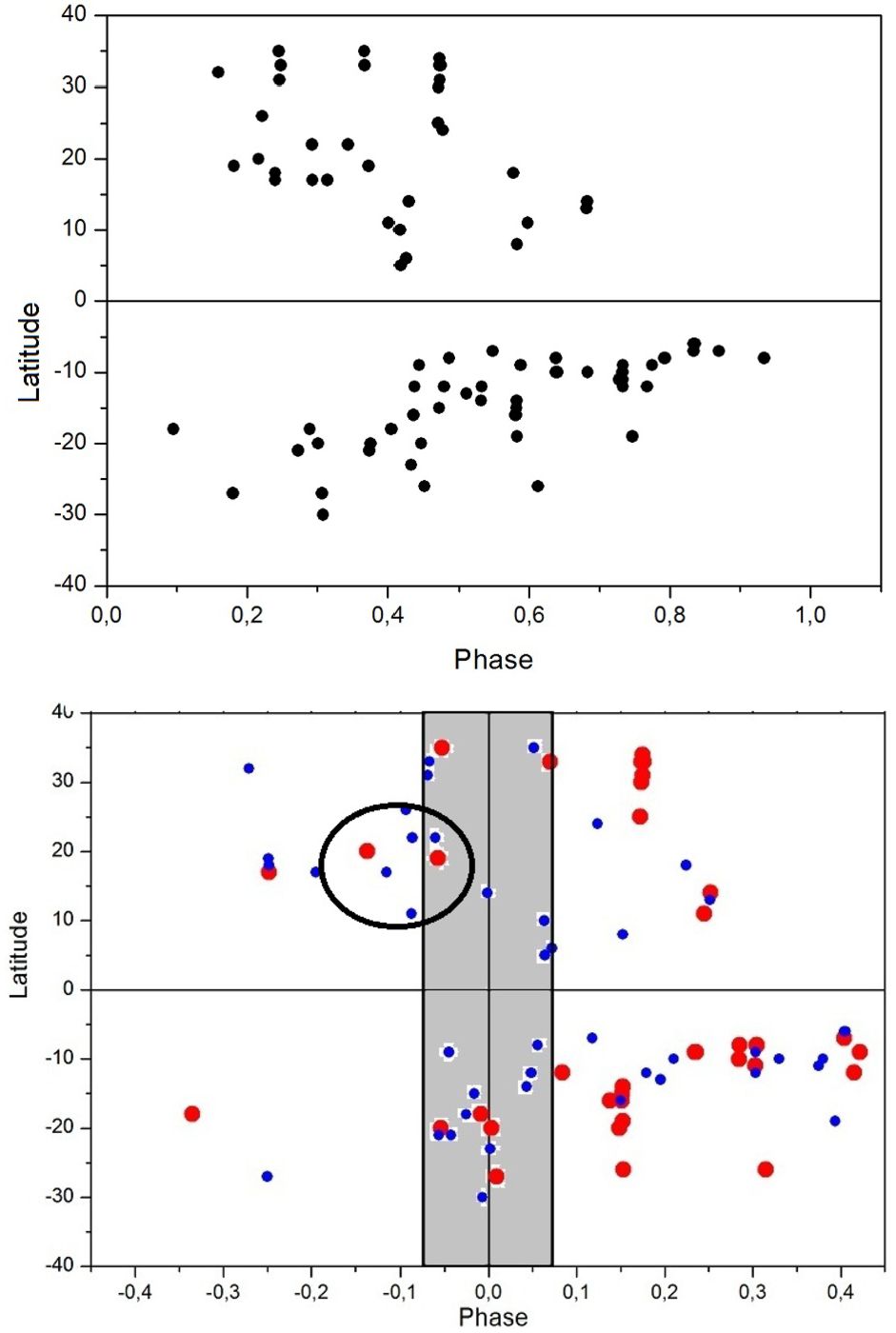}
\end{center}
\small
        \caption{Synthetic time-latitude diagram for the greatest X-ray flares. Top -- beginning of the cycles are identified, both kinds of lares are given as black points. Bottom -- the cycles maxima are identified, vertical black line corresponds to the cycles maxima, the vertical gray strip shows the phase of maximum, red points are for the flares with magnitudes larger then X10, blue ones stand for the magnitudes between X4 and X10.  
       }
\label{F3}
\end{figure}

The number density of the points on the  plots is, obviously, insufficient to draw a final conclusion. In order to improve the situation, we use the method of superimposed epochs (Fig.~\ref{F3}). Note that in the southern hemisphere, the pattern is more intensive and long than in the northern one.

Fig.~\ref{F3} illustrates two options for constructing the synthetic diagram: one by identifying the onset of the cycles (upper panel) and the other, by identifying the cycle maxima (lower panel). The second option seems to be more informative, and we consider it as the basic one. We see here two patterns in each hemisphere. The more intensive wave propagates towards the equator in the right part of the panel, while the other, less intensive, propagates towards the pole in the left part of the panel.  Both waves collide in the area shown with an ellipse in Figs. 3-4,   further referred to as a burning point of the wave of flares.

\begin{figure}    
\begin{center}
\includegraphics[width=0.85\textwidth]{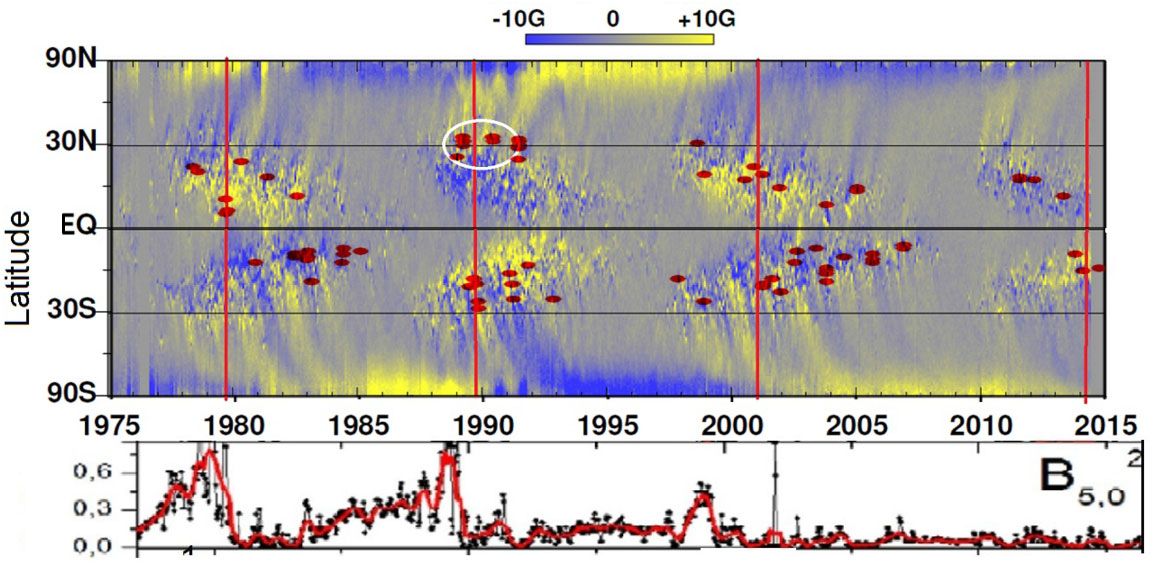}
\end{center}
\small
        \caption{A burning point of the wave of flares (the white ellipse on the top panel) against the surface magnetic field. On the top panel, this position is compared with space-time distribution of the surface magnetic field from \cite{H15}(shown in colors). The red lines mark the dates of the cycle maxima. On the lower panel, the location of the burning point is compared with the evolution of 5th harmonics of the surface solar magnetic field. 
       }
\label{F4}
\end{figure}

The position of the burning point in the course of an activity cycle is shown in Fig.~\ref{F4} against the surface solar magnetic field. We see that at the burning point, the magnetic waves propagating towards the equator (the main activity wave) and the wave propagating to the pole separate. Thus, the burning point means the start of intensive flare activity. It exists in both cycles in both hemispheres. To avoid overloading the figure, we have shown only one of the white ellipses in Fig. 4 in the northern hemisphere near the 1990 maximum. This point agrees both spatially and essentially with another concept, the overlapping point (OLP), which was introduced by \cite{Oetal23}.

This paper devoted to the analysis of extended cycles shows that 1-2 years before the maximum of each cycle, three waves coexist on the Sun in each hemisphere: the wave of the previous cycle and two waves of the new cycle (see Fig. 4, upper panel). In the new cycle, one wave moves to the pole and the other, to the equator. It was proposed to call this point the {\it overlapping point}  (OLP). Due to the presence of this point, the fifth harmonic is significantly amplified when the magnetic field is decomposed into micropoles (see Fig. 4, lower panel). It is important to note that it is here, at the end of the growth phase of the cycle, that the first X-ray flares occur. The lower panel shows the evolution of the harmonic $B_5$ (the data are taken from \cite{Oetal23}), which corresponds to the $l=5$ in the magnetic field decomposition. 

Thus, the concepts of burning point and overlapping point (OLP) coincide in time and space, although they reflect different aspects of similar processes in fields of different scales.

\begin{figure}    
\includegraphics[width=0.9\textwidth]{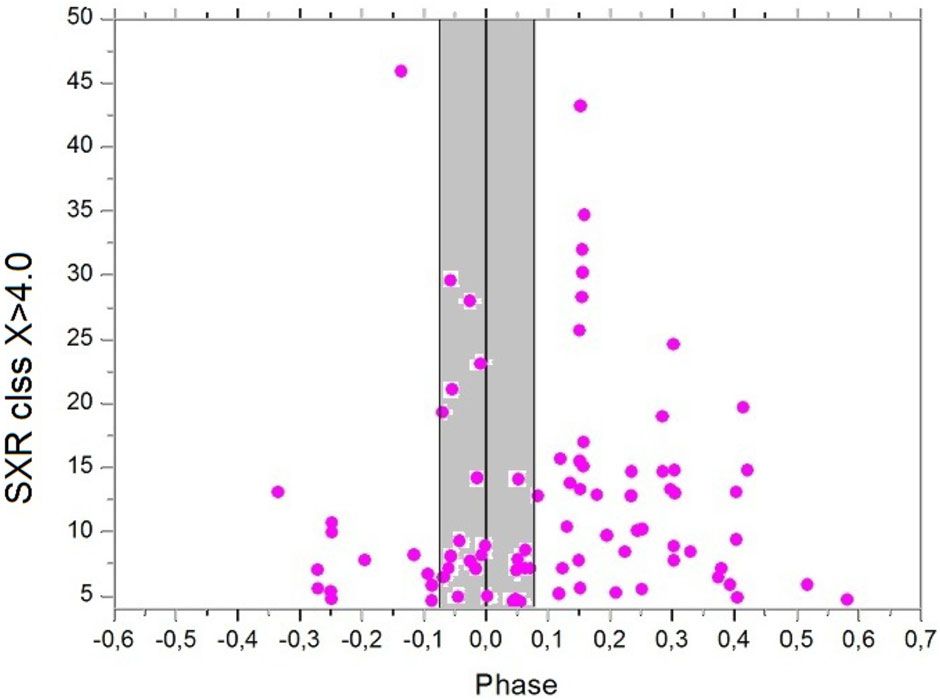}
\small
        \caption{Relation between the SXR class and the phase of the activity cycle. The vertical line shows the cycle maximum, the gray vertical strip stands for the phase of maximum. The dots indicate flares. }
\label{F5}
\end{figure}

Fig.~\ref{F5} shows the time and the peak intensity of the observed X-ray flares. The letter X is omitted, and the scale along the ordinate axis is given in units of $10^{-4}$ W/m$^2$. As in Figure 3b, the abscissa indicates the observation time as a fraction of the cycle length. As follows from Figs.~\ref{F4}, \ref{F5}, the burning starts at the relative high latitudes, near the point where equatorward and polarward waves coincide.   In the growth phase, X-ray flares are virtually absent. Then, 1-2 years before the maximum, relatively weak flares of magnitude X4-X7 appear, and subsequently, such flares occur throughout all phases of the cycle. Powerful flares actually avoid the growth phase. They appear at the cycle maximum and, then, occur all along the decay phase. The most powerful flares occur at the highest latitudes at the beginning of the decay phase. At this time, an interaction of two ways still persist.

As we noted earlier, the occurrence rate of flares changes depending on the intensity of the event. \cite{KOSL2022} showed that low-energy flares of classes B and C do not depend much on spottedness. This dependence increases with the level of spottedness and the phase of the cycle. In Figs 3 and 5, the zone of the cycle maximum is shaded gray. Of course, the beginning and the end of the maximum phase are weakly defined. \cite{ISH2022} suggests that the maximum phase starts at the point where the spottedness reaches 85\%\ of the expected height of the  maximum. Unfortunately, the exact value can only be determined later, and its physical basis is not clear. The time interval in the figure covers approximately $\pm$8 months relative to the calendar date of the maximum. This agrees fairly well with the SSN curve on the bottom panel of Fig. 1. It is easy to see that the most powerful flares are located near the maximum and especially during the decline phase.

Thus, of all flares of class X4-10 (Table 1, excluding data for cycle 25), 11 events occurred in the growth phase, 18 events at the maximum, and 20 events in the decay phase. For powerful events X>10, the ratio is even more striking: only 2 events occurred in the growth phase, 8 at the maximum, and 27 in the decline phase. It can be argued that powerful events avoid growth phases and appear first immeadiately near the maximum at the overlapping point (OLP).

\section{Conclusions: The largest flares in the general concept of the solar cycle}

To summarize, we conclude that the largest solar X-ray flares are an integral part of the solar activity cycle understood as a complex phenomenon that includes various particular details, such as sunspots, active regions, etc. This becomes particularly clear when we consider the most powerful flares, while the weak flares occur more or less at any time.

The largest flares occur in each hemisphere mainly in two waves of activity. One of these propagates towards the equator and is similar to the conventional activity waves observed in sunspot data. The other (less intensive) propagates towards the pole. The waves meet in the vicinity of a point on the time-latitude diagram which separates the activity waves of the surface magnetic field. In other words, the interaction of the local and global magnetic fields in this area increases the probability of formation of complicated magnetic structures and the occurrence of the largest X-ray flares.     

The point is that one cannot predict the magnitude of the flare considering only the intrinsic structure of the magnetic field inside the active region. Both magnitude and geoeffectiveness are determined by complicated interplay between the active region and the surrounding medium including the coronal magnetic field and coronal holes, nearby active regions, and global magnetic topology of the Sun.   

Of course, there are important papers where the authors consider specific flares and arrived at similar conclusions. In particular, \cite{Eetal24}
note that the proximity of the flare to a large coronal hole influences the expansion and propagation of the coronal mass ejection toward the Earth, probably, impacting the solar wind speed and density. Moreover, they note a sudden expansion of the coronal hole during the flare. 

\cite{KYSetal24} note that the geomagnetic storm of magnitude G5 was produced by a sequence of eruptions at the conjugation of two active regions, namely  AR13664 and AR13668. 

\cite{Setal23} stress the importance of complex magnetic and energy relationships between active regions and the corona on both local and global scales (see also \cite{MTetal25}).

\section{Discussion}

The question is, why do the largest X-ray flares occur mainly at the point where the activity waves overlap. First of all, note that X-ray flares are the very powerful ones.  Besides, the X-ray emission exists at any time. but due to different reasons (position on or behind the disc, instrumental sensitivity, etc.) it remains unobservable. 
It is generally accepted that the flare is the result of transformation of the magnetic energy into another form of energy. However, an important point is that part of the magnetic energy available for transformation is associated with electric currents, and the accumulation of this energy requires special conditions.

Note, that the equatorward waves are mainly the zonal structures of very high even order. These harmonics are related to the toroidal configuration. The sunspot groups are mainly simple bipolar structures with toroidal magnetic lines following the classical  Hale polarity law. Here, the magnetic lines  are closed inside the activity area and do not propagate very high. However, the magnetic structure at the periphery is very  different. Magnetic lines go primarily in the meridian direction, their height is much greater, and they extend to the corona creating very high loops. The loops can be considered a global phenomenon. The sign of the magnetic field is opposite to that of the active region.  

The propagation of the activity wave discussed above is compatible with solar dynamo theory. Indeed, the wave vector $\bf k$ of the activity wave obtained within the framework of the short-wavelength approximation for the \cite{P55} migratory dynamo leads us to the following governing equation (e.g., \cite{KS95}).

\begin{equation}
    (\Gamma +k^2)^2-i \alpha {\bf k} \cos \theta =0\,,
\end{equation}
where $\Gamma$ is the complex growth rate, $\alpha$ stands for distribution of the dynamo drivers, $\theta$ is the latitude, and $\cos\theta$ represents the curvature effects. This is a fourth-order algebraic equation with four roots ${\bf k}_i(\theta)$, $i=1 \dots 4$. Two roots, $i=1,2$ are required to be combined in the main equatorwards activity wave while two other roots $i = 3,4$  describe the secondary wave propagating polewards. Imprints of this secondary wave can be found in numerical solar dynamo models starting from \cite{IR77}.

The general conclusion from the comparisons above is that powerful X-ray flares avoid to appear in the growth phase of the cycle. This is especially noticeable in the case of the most powerful flares with $>$X10. The first appearance of powerful X-ray flares is recorded shortly before the maximum and coincides with a specific structure of the large-scale field.  
Subsequently, X-ray flares appear on the time-latitude diagram, usually on the line of interaction of large-scale and local fields.
This means that the magnetic field is involved in generation of large flares not only in the active region, but also in the entire activity complex including one or more active regions and adjacent structures of the large-scale field (in particular, coronal holes) \cite{OS13}. 

Consider as an example the situation in october-november 2003. AR 10486 produced a large number of powerful flares (“halloween" series). The region passed through the central meridian on 29.10.2003 at the latitude S19. In the vicinity of the main active region, a few other large AR and an equatorial hole (see Fig.6) were observed to form a single activity complex.

This structure is characteristic of global complexes of activity.
The concept has been discussed by different authors for nearly 90 years since the publication of M.N.Gnevyshev, who called it “impulses of solar activity” \cite{G38} (the English version of the paper is available in ADS (Astrophysics data system) as \cite{G38a}). In this and the following publications, the concept did not include elements of the large-scale field. A more precise definition was given by \cite{OS13}. A global activity complex (GAC) is defined as the appearance of one or several closely located, developed active regions with an adjacent zone of open magnetic field lines in the form of a coronal hole. As a rule, GACs penetrate high into the corona and are accompanied by significant flare activity and the occurrence of coronal mass ejections (CMEs), which are associated with the most powerful flares. This explains why X-ray flares virtually avoid the growth phase of the 11-year cycle, appearing only near the peak and, then, throughout the decline phase. The fact is that such events are associated with GACs, which, in turn, combine the properties of local and large-scale magnetic fields, i.e., active regions and coronal holes. The latitudinal distribution of coronal holes directly depends on the phase of the 11-year cycle. At the minimum of the cycle, the coronal holes dominate, while the equatorial holes are extremely rare. As we approach the peak of the cycle, the number of sunspots and active regions increases. At the same time, the number of equatorial coronal holes also begins to increase. Then, in the course of evolution of the cycle, the latitudinal distribution of equatorial coronal holes broadens. It reaches its peak after the activity maximum and, then, begins to decay. In other words, it is at the decay branch of the cycle that we observe the most number of equatorial coronal holes.

The appearance of coronal holes is associated with the configuration of the solar magnetic field (\cite{Letal97, Hetal20}). Coronal holes are the areas where magnetic field lines are open, allowing plasma streams to escape freely into space. During the declining phase of the solar cycle, the so-called equatorial dipole often predominates. This large-scale magnetic field structure is responsible for the appearance of low-latitude (equatorial) coronal holes. The transition of the axial dipole to the equatorial one and back and the connection with the phases of the solar cycle is described in detail by \cite{LO06, Oetal20}.

\begin{figure}    
\includegraphics[width=0.9\textwidth]{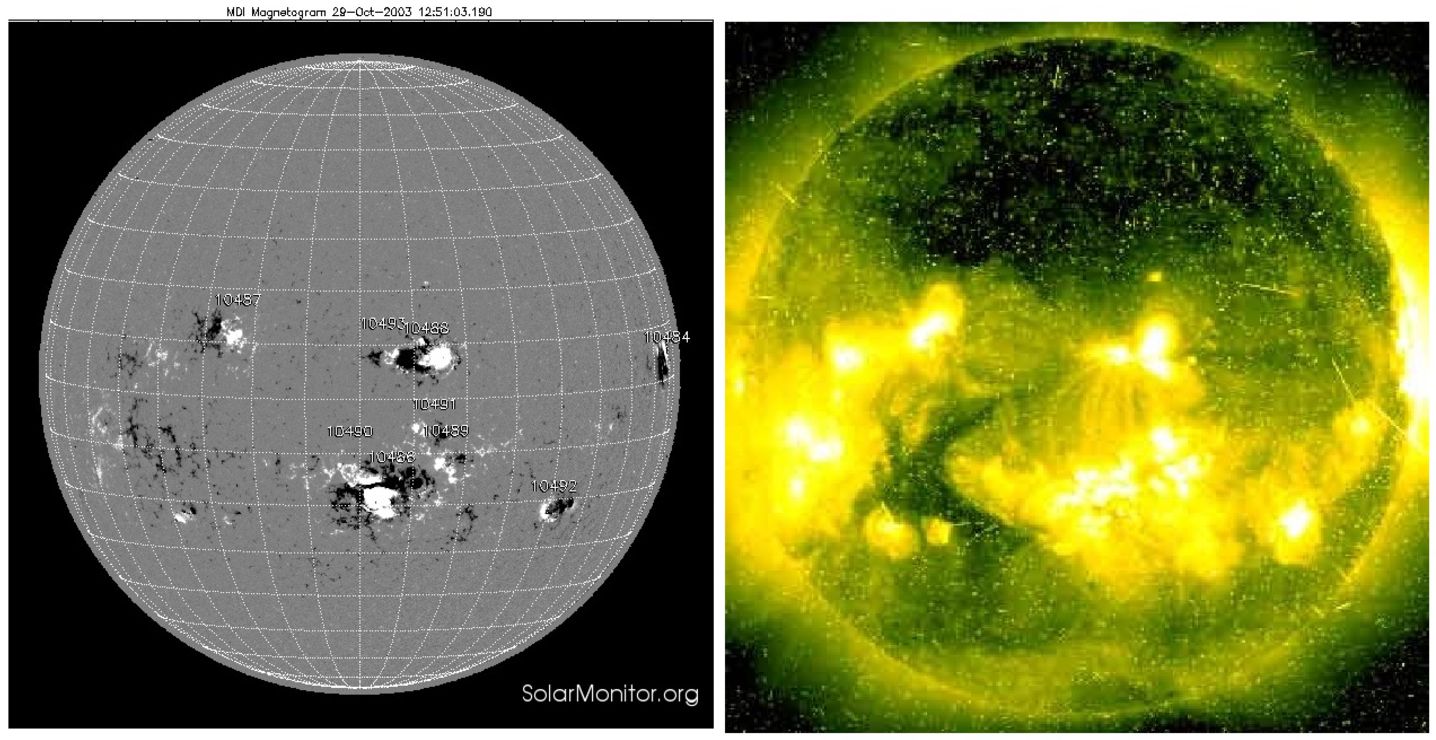}
\small
        \caption{Solar activity on October 29, 2003. Left - the magnetogram; right - the image in $\lambda 284 A.$}
\label{F6}
\end{figure}

 The interaction of loops and active regions affects the simple bipolar structure creating complex groups. The neutral lines separating the structures are much more complicated than a simple neutral line separating bipolar structures. The interaction of structures of both types creates currents and ensures rapid propagation 
 and other properties typical of the powerful active areas producing the largest X-ray flares.  The above points are mentioned by \cite{BO69, OS13, Betal00, Z24}
 and, in particular, in connection with the properties of the heliospheric equator (see, e.g., \cite{O10, Oetal11}).

The topic under discussion is directly related to the analysis of superflares on the late-type stars by \cite{Ketal99, LL02, Letal03}.
  
All authors  equally contributed to the manuscript.

The study was conducted under the state assignment of IZMIRAN and Lomonosov Moscow State University.

All data on photospheric magnetic fields are available at the\\ 
\url{http://wso.stanford.edu/synopticl.html}
and  Sunspot data from the World Data Center SILSO, Royal Observatory of Belgium, Brussels. Version V2. 

The authors declare that they have no conflict of interest.


\phantomsection
\bibliographystyle{rusnat} 
\bibliography{bibliography}


\end{document}